# Facet-dependent magnon-polarons in epitaxial ferrimagnetic $Fe_3O_4$ thin films

Wenyu Xing[1*], Yang Ma[1], Yunyan Yao[1], Ranran Cai[1], Yuan Ji[1], Richen Xiong[1], Ka Shen[2], and Wei Han[1,3*]

[1] International Center for Quantum Materials, School of Physics, Peking University, Beijing, 100871, P. R. China

[2] The Center for Advanced Quantum Studies and Department of Physics, Beijing Normal University, Beijing 100875, P. R. China

[3] Collaborative Innovation Center of Quantum Matter, Beijing, 100871, P. R. China

*Correspondence to: wenyuxing@pku.edu.cn (W.X.) and weihan@pku.edu.cn (W.H.)

**Abstract**

Magnon-polarons are coherently mixed quasiparticles that originate from the strong magnetoelastic coupling of lattice vibrations and spin waves in magnetic-ordered materials. Recently, magnon-polarons have attracted a lot of attention since they provide a powerful tool to manipulate magnons, which is essential for magnon-based spintronic devices. In this work, we report the experimental observation of facet-dependent magnon-polarons in epitaxial ferrimagnetic $Fe_3O_4$ thin films via spin Seebeck effect measurement. The critical magnetic fields for the magnon-polarons in the (110)- and (100)-oriented $Fe_3O_4$ films are 1.5 T and 1.8 T, respectively, which arises from the different phonon velocities along the [110] and [100] directions. As the temperature decreases, the magnon-polarons-enhanced spin Seebeck voltage decreases in both (110)- and (100)-oriented $Fe_3O_4$ films, which could be attributed to the enhanced magnon-polarons scattering at elevated temperatures. This work demonstrates the crystal structure engineering in epitaxial magnetic films as a promising route to manipulate the magnon-polarons for future magnon spintronic applications.

## I. INTRODUCTION

As the data carriers, the properties of magnons are crucial for magnon-based spintronic applications [1-5]. Magnon-polarons provide a powerful tool to improve the properties of magnons, such as increasing the lifetime and group velocity [6-10]. Magnon-polarons could be formed in magnetic-ordered materials due to the strong magnetoelastic coupling between spin waves and lattice vibrations. Recently, a magnon-polarons-induced anomaly has been discovered in the spin Seebeck effect (SSE) of yttrium iron garnet ($Y_3Fe_5O_{12}$: YIG) [7,8,11] and the spin Peltier effect of lutetium iron garnet [12]. Besides YIG, $Fe_3O_4$ is a ferrimagnetic insulator with strong magnon-phonon coupling [13,14], which makes it a potentially suitable platform to investigate magnon-polarons. More interestingly, neutron scattering studies have shown the anisotropic properties of phonons in $Fe_3O_4$ [15]. Thus, $Fe_3O_4$ might reveal the anisotropic properties of magnon-polarons which are expected to exist theoretically [8], but have never been experimentally explored yet.

In this work, we report the magnon-polarons in high-quality (110)- and (100)-oriented epitaxial $Fe_3O_4$ thin films *via* SSE measurements. Interestingly, the critical magnetic fields of magnon-polarons are found to be strongly facet-dependent in the (110)- and (100)-oriented $Fe_3O_4$ films, which is due to the anisotropic phonon velocities along the $Fe_3O_4$ crystal's [110] and [100] directions. The magnon-polarons-enhanced SSE voltages decrease in both (110)- and (100)-oriented $Fe_3O_4$ films as temperature increases, which could be attributed to the enhanced magnon-polarons scattering at elevated temperatures.

## II. EXPERIMENTAL

The $Fe_3O_4$ thin films were grown on (110)- and (100)-oriented MgO substrates (from Hefei Kejing Materials Technology) using oxide molecular-beam epitaxy system (MBE-Komponenten GmbH; Octoplus 400) with a base pressure lower than $1.0 \times 10^{-10}$ mbar [16]. Prior to the $Fe_3O_4$ film growth, the MgO substrates were pre-cleaned by annealing at 600 ℃ for 2 hours. Then, the $Fe_3O_4$ thin films were grown by evaporating Fe from a thermal effusion cell with a deposition rate of 0.02 Å/s in the diluted ozone gas under the pressure of ~ $5.2 \times 10^{-7}$ mbar. During the $Fe_3O_4$ film growth, *in situ* reflective high-energy electron diffraction (RHEED) was used to monitor the growth and to characterize the crystalline properties. Figures 1(a) and 1(c) show the typical RHEED patterns of the (110)-oriented MgO substrates and 100 nm thick (110)-oriented $Fe_3O_4$ thin films viewed from the MgO crystal's $[1\bar{1}0]$ direction. The sharp RHEED pattern indicates the high-quality crystalline properties of (110)-oriented $Fe_3O_4$ thin films epitaxially grown on the MgO substrates. Following the same growth procedures, high quality (100)-oriented $Fe_3O_4$ thin films were achieved on (100)-oriented MgO substrates (Figs. 1(b) and 1(d)). Figures 1(e) and 1(f) show the X-ray diffraction (XRD) results of the $Fe_3O_4$ films, which clearly show the epitaxial features of the (110)- and (100)-oriented $Fe_3O_4$ films on the corresponding MgO substrates. Furthermore, the Verwey transitions are observed for (110)- and (100)-$Fe_3O_4$ thin films (Figs. 1(g) and 1(h)), which are consistent with the previous reports and further confirm the $Fe_3O_4$ phase of our samples [17-19].

The $Fe_3O_4$ longitudinal SSE devices were fabricated using standard E-beam lithography and lift-off processes. The first step was to define a Pt electrode (width: 200 nm) on the $Fe_3O_4$ films *via* electron-beam lithography. The 10 nm thick Pt was deposited in a magneton sputtering system with a base pressure lower than $8.0\times10^{-7}$ mbar. Then a heater electrode consisting of 100 nm thick $Al_2O_3$ and 20 nm thick Ti was grown *via* E-beam evaporation. Figure 2(a) shows the schematic of the fabricated longitudinal SSE device with the measurement geometry, where the Ti layer was used to generate the heat flow from Joule heating, the Pt electrode was used to detect the SSE voltage and the $Al_2O_3$ layer was used to accomplish electric insulation between the Ti and Pt electrodes.

The magnon-polarons and SSE in $Fe_3O_4$ thin films were measured using on-chip local heating *via* standard low frequency lock-in technique in a Physical Properties Measurement System (PPMS, Quantum Design). A current source (K6221, Keithley) was used to provide the AC current (frequency: 7 Hz) in the Ti electrode to generate temperature gradient ($\nabla T$) perpendicular to the $Fe_3O_4$ films. The heating power was kept constant at 0.087 mW across the temperature range of measurement (2 K to 12.5 K). Additionally, the local temperature change in this study was a minor effect based on the analysis of the heating effect on the $Fe_3O_4$ films. The bottom surface of the SSE device was attached to the Cu sample holder using thermal conducting paste, where the Cu holder acted as a thermal sink. An in-plane magnetic field ($\mu_0 H$) was applied perpendicular to the Pt strip as shown in the right inset of Fig. 2(b). The SSE signals were measured *via* the second harmonic voltages using lock-in amplifiers (SR830, Stanford Research). During the measurement, a low noise voltage preamplifier (SR560, Stanford Research) was used to enhance the signal-to-noise ratio.

## III. RESULTS AND DISCUSSION

Figure 2(b) shows the representative SSE voltage curves of (110)-oriented $Fe_3O_4$ films as a function of magnetic field at $T$ = 2 K. At low magnetic fields, a hysteresis behavior of the SSE signal is observed between ± 0.3 T at $T$ = 2 K (left inset of Fig.

2(b)). The hysteresis loop of the SSE voltages follows the magnetization curves of the Fe$_3$O$_4$ as a function of the magnetic field. Interestingly, two enhancement anomalies of the SSE signals (indicated by the orange triangles) are observed clearly at $\mu_0 H \sim \pm$ 1.5 T, which are the typical experimental signatures of magnon-polarons in a magnetic material [7-9,20]. At the critical magnetic fields of $\mu_0 H \sim \pm$ 1.5 T, magnon-polarons form and enhance the SSE voltage signals as a result of the tangential magnon and phonon dispersions [8,20]. Theoretically, the magnon dispersion of ferrimagnetic Fe$_3$O$_4$ can be described by the following expression [7,8]:

$$\omega_k = \sqrt{D_{\text{ex}}k^2 + \gamma\mu_0 H}\sqrt{D_{\text{ex}}k^2 + \gamma\mu_0 H + \gamma\mu_0 M_s \sin^2\theta} \#(1)$$

where $\omega_k$ is the angular frequency of magnon, $k$ is the wave vector, $D_{\text{ex}}$ is the exchange stiffness coefficient, $\gamma$ is the gyromagnetic ratio, $\mu_0 H$ is the external magnetic field, $\mu_0 M_s$ is the saturation magnetization, and $\theta$ is the angle between the external magnetic field and the magnon propagation direction. The transverse-acoustic (TA) phonon dispersion can be described by the following expression [7,8]:

$$\omega_{\text{TA}} = c_\perp k \#(2)$$

where $\omega_{\text{TA}}$ is the angular frequency of TA-phonon, and $c_\perp$ is the TA-phonon velocity. The magnon and phonon dispersion curves of (110)-oriented Fe$_3$O$_4$ are plotted in Figs. 2(c-e). At the critical magnetic field, $\mu_0 H = \mu_0 H_{\text{TA}}^{[110]}$, the magnon and phonon dispersion curves of (110)-oriented Fe$_3$O$_4$ are tangential to each other (Fig. 2(d)), which results in the enhancement anomalies of the SSE signals (orange triangles in Fig. 2(a)). For a small magnetic field, $\mu_0 H < \mu_0 H_{TA}^{[110]}$, the magnon and phonon dispersion curves are intersecting (Fig. 2(c)). For a large magnetic field, $\mu_0 H > \mu_0 H_{TA}^{[110]}$, the magnon and phonon dispersion curves are detached (Fig. 2(e)). By using the typical experimental values of $D_{ex} \sim 6.76 \times 10^{-6}$ m$^2$/s [21], $\mu_0 M_s$ of 364 emu/cm$^3$ [22], and $c_\perp \sim 3240$ m/s for $\boldsymbol{k}$ // [110] direction [15], the critical magnetic field ($\mu_0 H_{TA}^{[110]}$) is estimated to be ~2 T. The calculated $\mu_0 H_{TA}^{[110]}$ is slightly higher than the experimental results, which might result from the slight modification of material parameters in our epitaxial thin films compared to their bulk values. Compared to the magnon-polarons-enhanced SSE signals observed in YIG [7,11], a much broader peak shape is observed in Fe$_3$O$_4$ thin films, which could be attributed to the large magnon-phonon coupling in Fe$_3$O$_4$ [8]. Notice that the high longitudinal-acoustic phonon velocity of 7600 m/s for Fe$_3$O$_4$ [15] could lead to a large critical magnetic field of ~ 12.5 T for the magnon-polarons-enhanced SSE signal, which is above the maximum magnetic field of 9 T in our PPMS system.

Next, the anisotropic properties of magnon-polarons in the (110)- and (100)-oriented Fe$_3$O$_4$ thin films are investigated. Figure 3(a) shows the results of SSE signals in Fe$_3$O$_4$ thin films as a function of the magnetic field at $T$ = 2 K. The critical magnetic field of the magnon-polarons is ~1.8 T for (100)-oriented Fe$_3$O$_4$, which is larger than that of (110)-oriented Fe$_3$O$_4$ (~1.5 T). In the following, we discuss the physical mechanisms that account for the facet-dependent critical magnetic fields in (110)- and (100)-oriented Fe$_3$O$_4$ thin films. The magnon dispersion curves have been shown to be isotropic in previous inelastic neutron scattering studies [14,23-25]. On the other hand, the TA-phonon velocities for **k** along [110] and [100] direction of Fe$_3$O$_4$ crystal are 3240 m/s and 3430 m/s, respectively [15]. Hence, the different phonon velocities could lead to the experimental observation of the smaller $\mu_0 H_{\text{TA}}^{[110]}$ and larger $\mu_0 H_{\text{TA}}^{[100]}$, as illustrated in Figs. 3(b-c).

Figures 4(a-b) show the magnetic field dependence of SSE voltages for (110)- and (100)-oriented Fe$_3$O$_4$ at $T$ = 2, 5, 7.5 and 10 K, respectively. Figure 4(c) summarizes the magnon-polarons-enhanced SSE voltages ($V_{\text{TA}}$) as a function of the temperature. As temperature increases, $V_{\text{TA}}$ exhibits a strong suppression and disappears around 10 K. Since the magnon-polarons-enhanced SSE arises from the much longer phonon lifetimes than the magnon lifetimes ($\tau_{\text{ph}}/\tau_{\text{mag}} > 1$), this observation could be attributed to the smaller ratio of $\tau_{\text{ph}}/\tau_{\text{mag}}$ as temperature increases. In other words, the enhancement of the magnon-polarons scattering rate at elevated temperatures leads to the reduction of the magnon-polarons-induced SSE anomaly [7-9]. The critical magnetic fields exhibit little variation as a function of the temperature, as shown in the inset of Fig. 4(c), which is expected in the small temperature range from 2 K to 10 K.

Based on these results, it can be concluded that the present study for the facet-dependent magnon-polarons in epitaxial ferrimagnetic Fe$_3$O$_4$ thin films is an interesting piece that complements the results for the magnon-polarons of YIG/Pt heterostructures [7,12] as well as the spin caloritronics effects of $Fe_3O_4$/Pt heterostructures [26,27]. These results demonstrate that the crystalline structure plays an important role in the magnon-polarons-induced anomalies, which could provide a powerful method for magnon control. Hence, it might be useful for the research field of magnon spintronics that utilizes magnon for information carriers.

## IV. CONCLUSIONS

In summary, the facet-dependent magnon-polarons in high-quality epitaxial $Fe_3O_4$ thin films are experimentally investigated. Different critical magnetic fields for the magnon-polarons are observed in (110)- and (100)-oriented $Fe_3O_4$ thin films, which can be well interpreted by the anisotropy of phonon velocities. Furthermore, the stronger magnon-polarons scattering at higher temperatures results in the decrease of magnon-polarons-enhanced SSE voltages. Our work emphasizes the function of crystal directions in the magnon-polarons-induced SSE anomalies, and enriches the field of the magnon-polarons-dominated spin caloritronics effects in magnetic insulators.

**Acknowledgement**

We acknowledge the financial support from National Key Research and Development Programs of China (2018YFA0305601), National Natural Science Foundation of China (11974025, 11974047), Beijing Natural Science Foundation (No.

1192009), and the Strategic Priority Research Program of the Chinese Academy of Sciences (Grant No. XDB28000000).

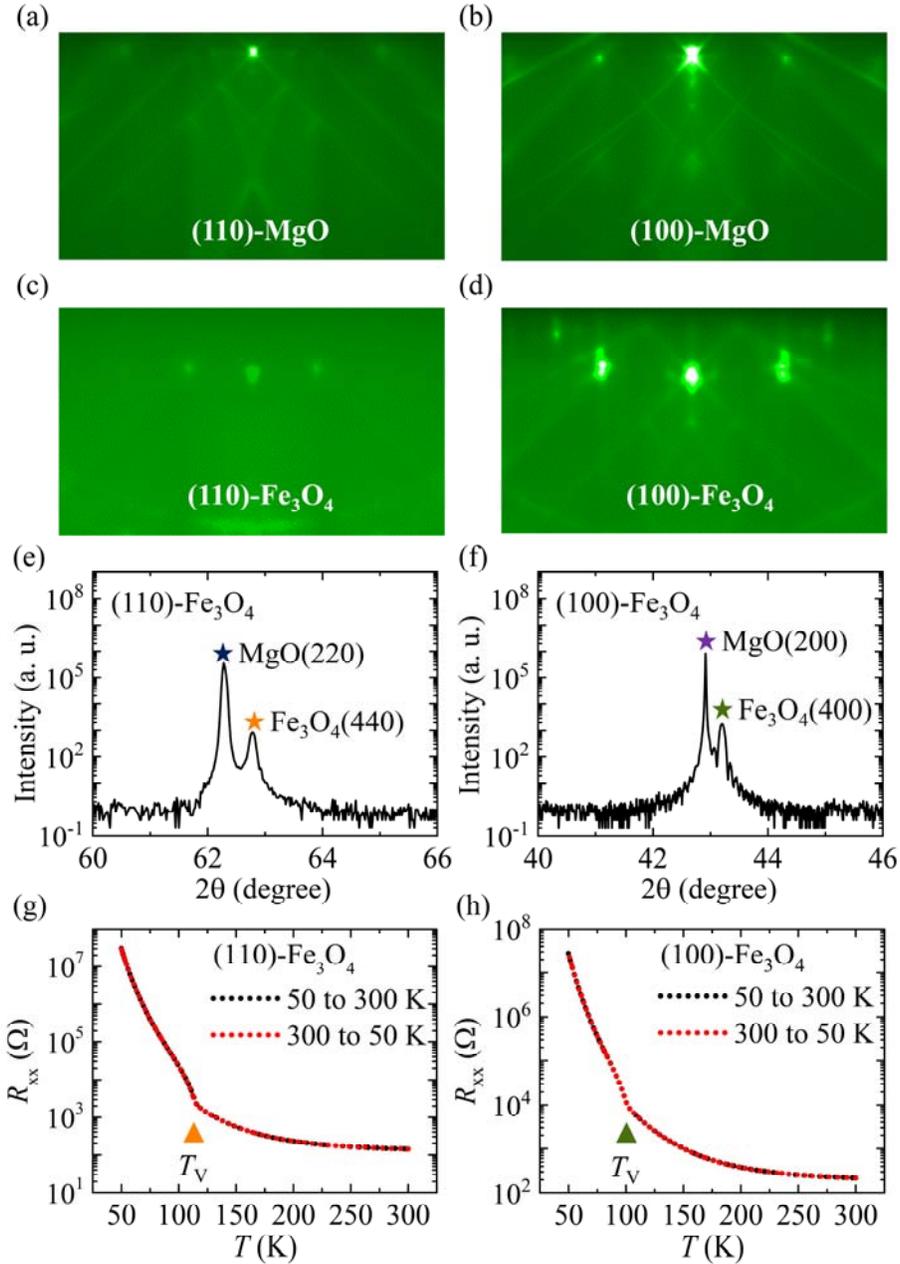

FIG. 1. Characterization of epitaxial Fe$_3$O$_4$ thin films. (a-b) RHEED patterns of the (110)- and (100)-oriented MgO substrates viewed from the crystalline [1$\bar{1}$0] and [001] directions, respectively. (c-d) RHEED patterns of the (110)- and (100)-oriented Fe$_3$O$_4$ thin films (thickness: 100 nm) grown on the corresponding MgO substrates. (e-f) XRD results of the Fe$_3$O$_4$ thin films. (g-h) The resistance as a function of temperature for the Fe$_3$O$_4$ thin films. $T_V$ represents the Verwey transition temperature.

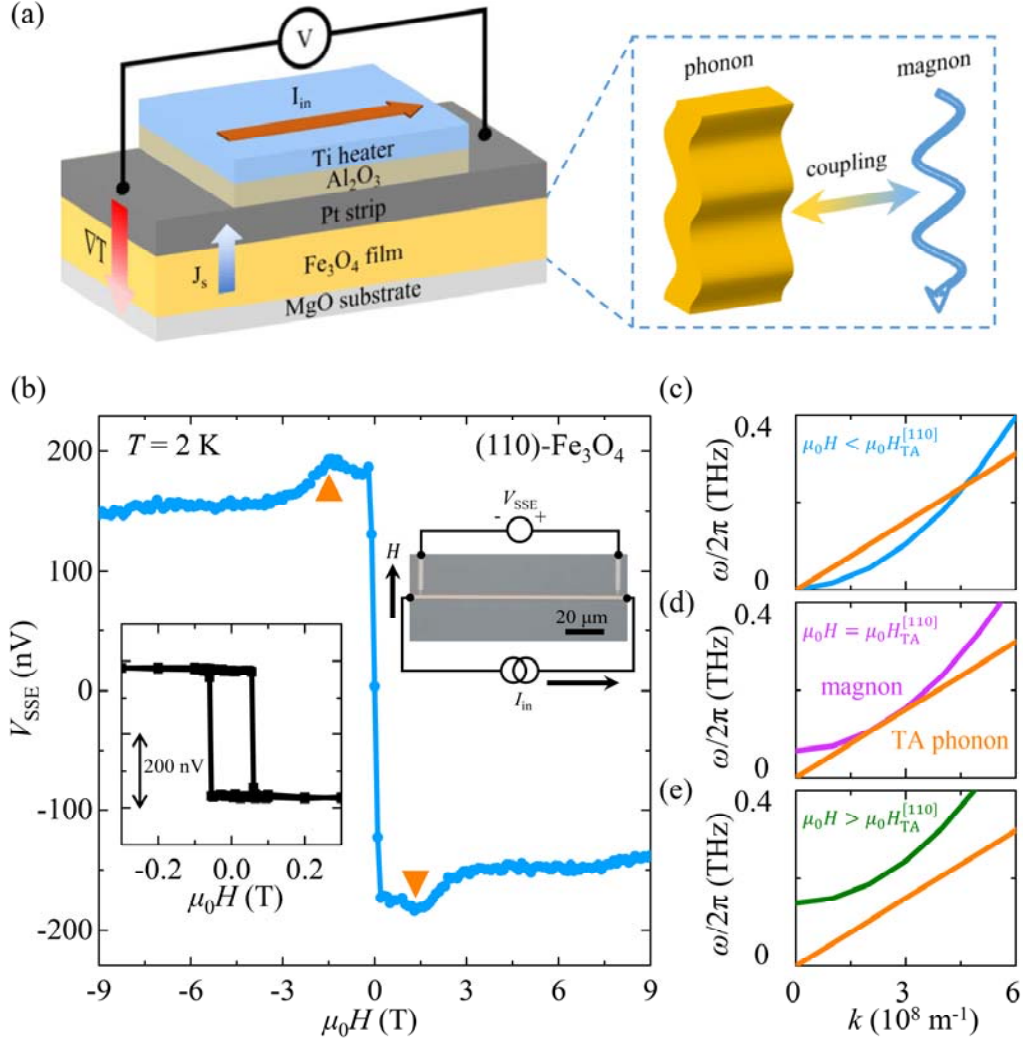

FIG. 2. Magnon-polarons in the epitaxial $Fe_3O_4$ thin films probed by SSE measurement. (a) Schematic of the $Fe_3O_4$ SSE device structure and the measurement geometry. The heater generates a temperature gradient ($\nabla T$) perpendicular to the film plane, resulting in the spin current ($J_s$) injection from $Fe_3O_4$ to the Pt electrode. The voltage meter measures the SSE signals. The right figure illustrates the coupling of lattice vibrations and magnons. (b) Experimental observation of magnon-polarons in the (110)-oriented $Fe_3O_4$ thin film (100 nm) at $T = 2$ K. The triangles indicate the formation of magnon-polarons. Bottom left inset: The hysteresis loop of the SSE voltages with the magnetic field between $\pm 0.3$ T. Top right inset: The optical image of the SSE device. (c-e) Magnon and TA-phonon dispersions with $k$ along the [110] direction under the magnetic fields of $\mu_0 H <$, $=$, and $> \mu_0 H_{TA}^{[110]}$, respectively.

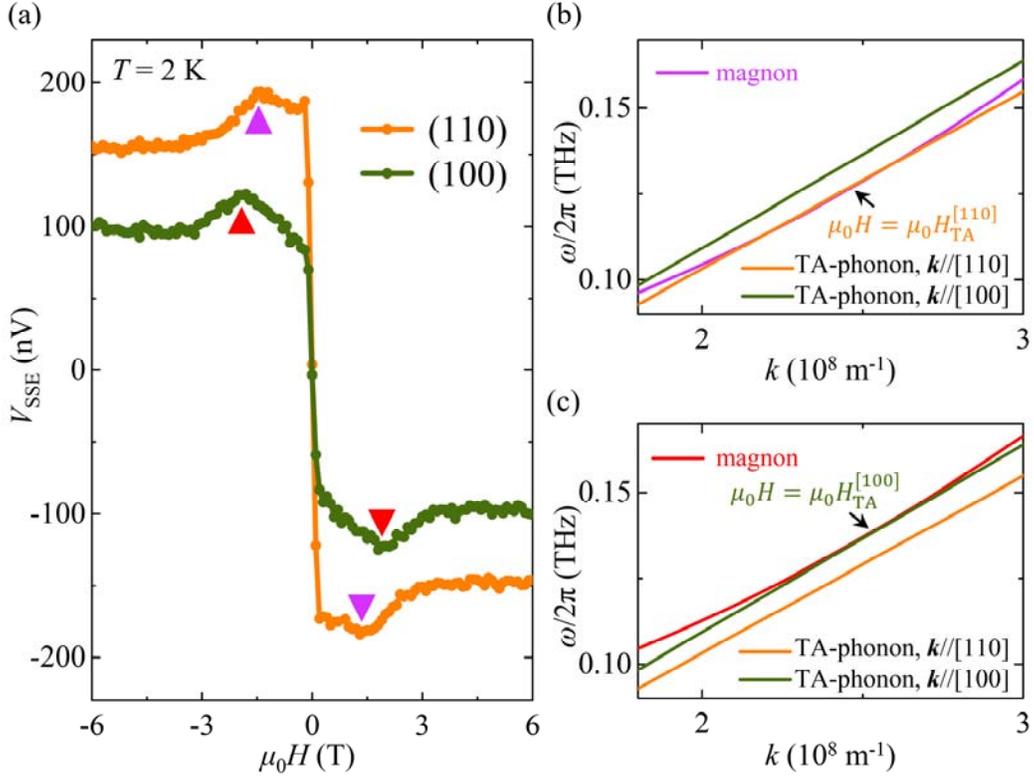

FIG. 3. Facet-dependent magnon-polarons in epitaxial $Fe_3O_4$ thin films. (a) The SSE voltages as a function of magnetic field for (110)- and (100)-oriented $Fe_3O_4$ thin films at $T = 2$ K. The triangles indicate the magnon-polarons-enhanced SSE signals. (b-c) Illustration for magnon and phonon dispersions at $\mu_0 H = \mu_0 H_{TA}^{[110]}$ and $\mu_0 H_{TA}^{[100]}$. The magnon dispersion is tangential to the TA-phonon dispersions with $k$//[110] and $k$//[100] at $\mu_0 H_{TA}^{[110]}$ and $\mu_0 H_{TA}^{[100]}$, respectively.

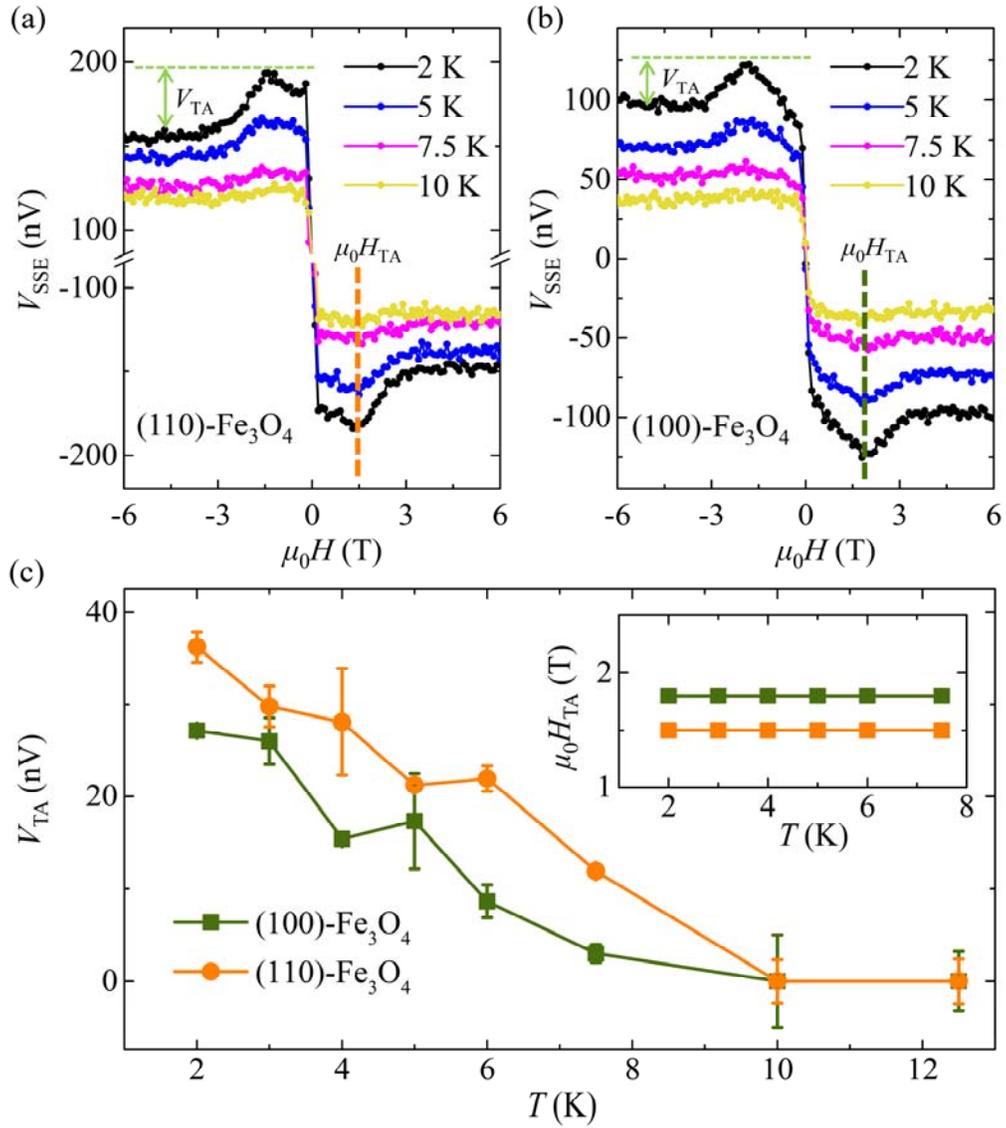

FIG. 4. Temperature dependence of magnon-polarons in epitaxial $Fe_3O_4$ thin films. (a-b) The magnon-polarons signals of (110)- and (100)-oriented $Fe_3O_4$ thin films at $T$ = 2, 5, 7.5, and 10 K, respectively. (c) The temperature dependence of the magnon-polarons-enhanced SSE voltages ($V_{TA}$) of (110)- and (100)-oriented $Fe_3O_4$ thin films. Inset: The critical magnetic fields of magnon-polarons as a function of temperature for (110)- and (100)-oriented $Fe_3O_4$ thin films.